\documentclass[12pt,prb,aps]{revtex4}

\usepackage{amsmath}

\def\uma{\rm 1\!\!\hskip 1 pt l}
\def\<{\langle}
\def\>{\rangle}

\newcommand{\lb}{\langle}
\newcommand{\rb}{\rangle}

\begin{document}

\title{Three Methods for Calculating the Feynman Propagator}

\author{F. A. Barone}
\email{fabricio@if.ufrj.br}
\author{H. Boschi-Filho}
\email{boschi@if.ufrj.br}
\author{C. Farina}
\email{farina@if.ufrj.br}
\affiliation{Instituto de F\'\i sica - Universidade Federal do Rio
de Janeiro, Caixa Postal 68528 - CEP 21945-970, Rio de Janeiro,
Brazil}

\begin{abstract}
We present three methods for calculating the Feynman propagator for
the non-relativistic harmonic oscillator. The first method was
employed by Schwinger a half a century ago, but has rarely been
used in non-relativistic problems since. Also discussed is an
algebraic method and a path integral method so that the reader
can compare the advantages and disadvantages of each method. 
\end{abstract}

\maketitle

\section{Introduction}
 The purpose of this paper
is to discuss three methods for calculating the Feynman
propagator. The methods are applied to the harmonic oscillator so
that they can be compared by the reader. The harmonic oscillator
was chosen because of its intrinsic interest and because it is the
simplest non-trivial system after the free particle (see for instance
references \onlinecite{Holstein98,BaroneFarina00}).

The first method we will discuss was developed by
Schwinger\cite{Schwinger51} to treat effective actions in quantum
electrodynamics and is based on the solution of the Heisenberg
operator equations of motion. 
The use of proper operator ordering and the subsidiary and
initial conditions yields the propagator. The second method is
based on algebraic techniques based on factorizing the time
evolution operator using the Baker-Campbell-Hausdorff
formulae.\cite{Wilcox,Merzbacher} By using factorization, the
completeness relations, and the value of the matrix element $\lb x
|p\rb$, we can determine the propagator. This method is close to
the one presented in Refs.~\onlinecite{Wang} and
\onlinecite{Yonei}, but here we will use the
Baker-Campbell-Hausdorff formulae in a slightly different way. 
The third method is a path
integral calculation that is based on a recurrence relation for
the product of infinitesimal propagators. As far as we know, this
recurrence relation has not appeared in previous discussions of
the one-dimensional harmonic oscillator path integral, although it
is inspired by a similar relation in the three-dimensional
system.\cite{Inomata} 

To establish our notation, we write the Feynman
propagator for a time independent non-relativistic system with
Hamiltonian operator
$\hat H$ in the form:
\begin{equation}
\label{propagadorFeynman}
K(x'',x';\tau)=\theta(\tau)\lb x''|\hat U(\tau)|x' \rb ,
\end{equation}
where $\hat U(\tau)$ is the time evolution operator:
\begin{equation}
\label{evolution}
\hat U(\tau)=\exp(-i \hat H\tau/\hbar),
\end{equation}
and $\theta(\tau)$ is the step function defined by:
\begin{equation}
\theta(\tau)= 
\begin{cases}
1 & \text{if $\tau \geq 0$} \\
0 & \text{if $\tau<0$}
\end{cases}
\end{equation}

\section{Schwinger's Method}\label{schwinger}
This method was introduced in 1951 by Schwinger in the context 
of relativistic quantum field theory\cite{Schwinger51} and it has
since been employed mainly in relativistic problems, such as the
calculation of bosonic\cite{Dodonov1} and
fermionic\cite{Dodonov2,Dodonov3,Likken91,BFV95} Green's
functions in external fields. However, this powerful method is
also well suited to non-relativistic problems, although it has
rarely been used in the calculation of non-relativistic Feynman
propagators. Recently, this subject has been discussed in an
elegant way by using the quantum action
principle.\cite{Schwinger2001} Before this time, only a few papers
had used Schwinger's method in this
context.\cite{Horing86,Farina93,FarinaRabello95} We adopt here a
simpler approach that we think is better suited for students and
teachers. 

First, observe that for $\tau>0$, Eq.~(\ref{propagadorFeynman}) 
leads to the differential equation for the Feynman propagator:
\begin{equation}
\label{eqdiffpropSchwing1}
i\hbar\frac{\partial}{\partial\tau}K(x'',x';\tau)=\lb x''|\hat
H\exp\bigl(-\frac{i}{\hbar}\hat H\tau\bigr)|x' \rb \,.
\end{equation}
By using the general relation between operators in the
Heisenberg and Schr\"odinger pictures, 
\begin{equation}
\label{relO_{H}->O_{S}Schwin}
\hat O_{H}(t)=e^{i \hat H t/\hbar}
\hat O_{S}e^{-i\hat H t/\hbar}\,,
\end{equation}
it is not difficult to show that if $\vert x\rangle$ is an 
eigenvector of the operator ${\hat X}$ with eigenvalue $x$, then it 
is also true that
\begin{equation}
{\hat X}(t)\,|x,t\rb =x|x,t\rb ,
\end{equation}
where
\begin{equation}
{\hat X}(t)=e^{i\hat H t/\hbar} {\hat
X} e^{-i \hat H t/\hbar}\,,
\end{equation}
and $\vert x,t\rangle$ is defined as
\begin{equation}
|x,t\rb =e^{i \hat H t/\hbar}
|x \rb \, .
\end{equation}
Using this notation, the Feynman propagator can be written as:
\begin{equation}
K(x'',x';\tau)=\langle x'',\tau|x',0\rangle\, ,
\end{equation}
where
\begin{subequations}
\label{ketsx''taux'0}
\begin{eqnarray}
\hat X(\tau)\vert x'',\tau\rangle&=&x''\vert x'',
\tau\rangle \\
\hat X(0)\vert x',0\rangle&=&x'\vert x',0\rangle
\end{eqnarray}
\end{subequations}
The differential equation for the Feynman propagator, 
Eq.~(\ref{eqdiffpropSchwing1}), takes the form
\begin{equation}
\label{propagadorconveniente}
i\hbar\frac{\partial}{\partial\tau}\langle x'',\tau|x',0\rangle
=\langle x'',\tau| \hat
H |x',0\rangle \,. \qquad (\tau>0)
\end{equation}

The form of Eq.~(\ref{propagadorconveniente}) is very suggestive
and is the starting point for the very elegant operator method
introduced by Schwinger.\cite{Schwinger51} The main idea is to
calculate the matrix element on the right-hand side of
Eq.~(\ref{propagadorconveniente}) by writing $\hat H$ in terms of
the operators 
${\hat X}(\tau)$ and ${\hat X}(0)$, appropriately ordered. 
Schwinger's method can be summarized by the following steps:

\begin{enumerate}
\item Solve the Heisenberg equations for the 
operators ${\hat X}(\tau)$ and ${\hat P}(\tau)$, which are given
by:
\begin{equation}
\label{12}
i\hbar\frac{d}{ dt}{\hat X}(t)=[{\hat X}(t),\hat H]
\quad \mbox{and} \quad
i\hbar\frac{d}{dt}{\hat P}(t)=[{\hat P}(t),\hat H]\, .
\end{equation}
Equation~(\ref{12}) follows directly from
Eq.~(\ref{relO_{H}->O_{S}Schwin}).

\item Use the solutions obtained in step (1)
to rewrite the Hamiltonian operator $\hat H$ as a function 
of the operators ${\hat X}(0)$ and ${\hat X}(\tau)$ ordered 
in such a way that in each term of $\hat H$, the operator 
${\hat X}(\tau)$ must appear on the left side, while the
operator 
${\hat X}(0)$ must appear on the right side. 
This ordering can be done easily with the
help of the commutator 
$[{\hat X}(0),{\hat X}(\tau)]$ (see
Eq.~(\ref{comutatorX(0)X(tau)})). We shall refer to the
Hamiltonian operator written in this way as the ordered
Hamiltonian operator 
$\hat H_{\rm ord}({\hat X}(\tau),{\hat X}(0))$. 
After this ordering, the matrix element on the right-hand side of 
Eq.~(\ref{propagadorconveniente}) can be readily evaluated:
\begin{eqnarray}
\label{definicaoF}
\langle x'',\tau| \hat H |x',0\rangle &=&\langle x'',\tau| 
\hat H_{\rm ord}\biggl({\hat X}(\tau),{\hat
X}(0)\biggr)|x',0\rangle \nonumber \\
&\equiv& H(x'',x';\tau) \langle x'',\tau|x',0\rangle\,,
\end{eqnarray}
where we have defined the function $H$. The latter is a c-number
and not an operator. If we substitute this result in
Eq.~(\ref{propagadorconveniente}) and integrate over $\tau$, we
obtain:
\begin{equation}
\label{propagadorintegral}
\langle x'',\tau|x',0\rangle=C(x'',x')\exp\biggl(-\frac{i}{\hbar}
\int^{\tau} \! H(x'',x';\tau') d\tau'\biggr)\, ,
\end{equation}
where $C(x'',x')$ is an arbitrary integration constant.

\item The last step is devoted to the calculation of 
$C(x'',x')$. 
Its dependence on $x''$ and $x'$ can be determined by 
imposing the conditions:
\begin{subequations}
\label{15}
\begin{eqnarray}
\label{equivalente1}
\langle x'',\tau| {\hat P}(\tau)\,|x',0\rangle
 &=&-i\hbar\frac{\partial}{\partial x''}
\langle x'',\tau|x',0\rangle \\
\langle x'',\tau| {\hat P}(0) |x',0\rangle
&=&+i\hbar\frac{\partial}{\partial
x'}\langle x'',\tau|x',0\rangle\,. \label{equivalente2}
\end{eqnarray}
\end{subequations}
These equations come from the definitions in
Eq.~(\ref{ketsx''taux'0}) together with the assumption that the
usual commutation relations hold at any time:
\begin{equation}
[{\hat X}(\tau),{\hat P}(\tau)]=[{\hat X}(0),{\hat
P}(0)]=i\hbar\,.
\end{equation}
After using Eq.~(\ref{15}), 
there is still a multiplicative factor to be determined in
$C(x'',x')$. This can be done simply by imposing the propagator
initial condition:
\begin{equation}
\label{condicaoinicialSchwing}
\lim_{\tau\rightarrow 0^{+}}\langle x'',\tau|x',0\rangle
=\delta (x''-x')\,.
\end{equation}
\end{enumerate}

Now we are ready to apply this method to a large class of 
interesting problems. In particular, we shall calculate the
Feynman propagator for the harmonic oscillator.

The Hamiltonian operator for the harmonic oscillator can be written
as
\begin{eqnarray}
 \hat H = \frac{{\hat P}^{2}(\tau)}{2m}
+\frac{1}{2}m\omega^{2}{\hat X}^{2}(\tau)\,,
\label{Hosciladortau}
\end{eqnarray}
or
\begin{eqnarray}
\hat H = 
\frac{{\hat P}^{2}(0)}{2m}+\frac{1}{2}m\omega^{2}\hat
X^{2}(0)\,,
\label{Hoscilador}
\end{eqnarray}
because the Hamiltonian operator is time independent, 
despite the fact that the operators ${\hat P}(\tau)$ 
and ${\hat X}(\tau)$ are explicitly time dependent. 
It is matter of choice whether to work with the Hamiltonian
operator given by Eq.~(\ref{Hosciladortau}) or by
Eq.~(\ref{Hoscilador}). For simplicity, we choose the latter. 

As stated in step (1), we start by writing down the 
corresponding Heisenberg equations:
\begin{subequations}
\label{HeisenbergEquations}
\begin{eqnarray}
\frac {d}{dt}{\hat X}(t) &=& \frac{\hat P(t)}{m} \\
\frac {d}{dt}{\hat P}(t)&=&-m\omega^2{\hat
X}(t)\,,
\end{eqnarray}
\end{subequations}
whose solutions permit us to write for $t=\tau$ that:
\begin{equation}
\label{X(t)}
{\hat X}(\tau)={\hat X}(0)\cos\omega \tau 
+\frac{{\hat P}(0)}{m\omega}\sin\omega \tau\, .
\end{equation}
For later convenience, we also write the corresponding 
expression for ${\hat P}(\tau)$:
\begin{equation}
\label{P(t)}
{\hat P}(\tau)=-m\omega{\hat X}(0)\sin\omega\tau 
+{\hat P}(0)\cos\omega \tau\,.
\end{equation}
To complete step (2) we need to rewrite 
${\hat P}(0)$ in terms of ${\hat X}(\tau)$ and $\hat X(0)$, 
which can be done directly from Eq.~(\ref{X(t)}):
\begin{equation}
\label{P(0)}
{\hat P}(0)=\frac{m\omega}{\sin(\omega\tau)}\bigl[{\hat X}(\tau)
-{\hat X}(0)\cos\omega \tau
\bigr]\,.
\end{equation}
If we substitute this result into Eq.~(\ref{Hoscilador}), 
we obtain:
\begin{eqnarray}
\label{Hnaoordenado}
\hat H &=& \frac{m\omega^{2}}{2\sin^{2}(\omega\tau)}
\bigl[{\hat X}^{2}(\tau)+{\hat X}^{2}(0)\cos^{2}(\omega\tau) 
- {\hat X}(0){\hat X}(\tau)\cos(\omega\tau) \nonumber\\
&&{}-{\hat X}(\tau){\hat X}(0)
\cos(\omega\tau)\bigr]+\frac 12
m \omega^{2}{\hat X}^{2}(0)\,.
\end{eqnarray}
Note that the third term in Eq.~(\ref{Hnaoordenado}) is not
written in the appropriate order. By using the commutation
relation
\begin{eqnarray}
\label{comutatorX(0)X(tau)}
\bigl[{\hat X}(0),{\hat X}(\tau)\bigr]
&=&\bigl[{\hat X}(0),{\hat X}(0)\cos(\omega \tau) 
+\frac{{\hat P}(0)}{m\omega}\sin(\omega \tau)\bigr] \nonumber \\
&=& \frac{i\hbar}{m\omega}\sin(\omega\tau)\,,
\end{eqnarray}
it follows immediately that
\begin{equation}\label{X(0)X(tau)ordenado}
{\hat X}(0){\hat X}(\tau)={\hat X}(\tau){\hat X}(0)
+\frac{i\hbar}{m\omega}\sin\omega\tau\, .
\end{equation}
If we substitute Eq.~(\ref{X(0)X(tau)ordenado}) into 
Eq.~(\ref{Hnaoordenado}), we obtain the ordered Hamiltonian:
\begin{equation}
\hat H_{\rm ord}=\frac{m\omega^{2}}{2\sin^{2}(\omega\tau)}
\bigl[{\hat X}^{2}(\tau)+{\hat X}^{2}(0)
-2{\hat X}(\tau){\hat X}(0)\cos(\omega\tau)\bigr]
-\frac{i\hbar\omega}{2}\cot(\omega\tau)\,.
\end{equation}

Once the Hamiltonian operator is appropriately ordered, we can
find the function $H(x'',x';\tau)$ directly from its definition,
given by Eq.~(\ref{definicaoF}):
\begin{eqnarray}
\label{F(x'',x',tau}
H(x'',x',\tau)&=& \frac{\langle x'',\tau| \hat H |x',0\rangle}
{\langle x'',\tau|x',0\rangle} \nonumber \\
&=& \frac{m\omega^{2}}{2}
\bigl[({x''}^{2}+{x'}^{2})\csc^{2}(\omega\tau)-2x''x'\cot(\omega\tau)
\csc(\omega\tau)\bigr]-\frac{i\hbar\omega}{2}\cot(\omega\tau)\,.
\end{eqnarray}
By using Eq.~(\ref{propagadorintegral}), we can express the
propagator in the form:
\begin{eqnarray}
\label{29}
\langle x'',\tau|x',0\rangle 
&=& C(x'',x')\, 
\exp\biggl\{-\frac{i}{\hbar} \!
\int^{\tau}d\tau'\bigl[\frac{m\omega^{2}}{2}
\bigl(({x''}^{2}+{x'}^{2})\csc^{2}\omega\tau'
\nonumber \\
&&{} - 2x''x'\cot(\omega\tau)
\csc\omega\tau'\bigr)-\frac{i\hbar\omega}{2}
\cot\omega\tau'\bigr]\biggr\}.
\end{eqnarray}
The integration over $\tau'$ in Eq.~(\ref{29}) can be 
readily evaluated:
\begin{equation}
\label{propamenosdeC}
\langle x'',\tau|x',0\rangle=
\frac{C(x'',x')}{\sqrt{\sin(\omega\tau)}}
\exp\biggl\{\frac{im\omega}{2\hbar\sin(\omega\tau)}
\bigl[({x''}^{2}+{x'}^{2})\cos(\omega\tau)
-2x''x'\bigr]\biggr\}\,,
\end{equation}
where $C(x'',x')$ is an arbitrary integration constant to be 
determined according to step (3). 

The determination of $C(x'',x')$ is done with the aid of 
Eqs.~(\ref{15}) and
(\ref{condicaoinicialSchwing}). However, we need to rewrite the
operators ${\hat P}(0)$ and 
${\hat P}(\tau)$ in terms of the operators ${\hat X}(\tau)$ and 
${\hat X}(0)$, appropriately ordered. 
For ${\hat P}(0)$ this task has already been done 
(see Eq.~(\ref{P(0)})), and for ${\hat P}(\tau)$ we find after
substituting Eq.~(\ref{P(0)}) into Eq.~(\ref{P(t)}):
\begin{equation}
\label{P(tau)ordenado}
{\hat P}(\tau)=m\omega\cot(\omega\tau) \bigl[{\hat X}(\tau)-\hat
X(0)\cos\omega\tau\bigr]-m\omega{\hat X}(0)\sin(\omega\tau)\, .
\end{equation}
Then, by inserting Eqs.~(\ref{P(tau)ordenado}) and
(\ref{propamenosdeC}) into Eq.~(\ref{equivalente1}) it is not
difficult to show that:
\begin{equation}
\frac{\partial C(x'',x')}{\partial x''}=0\, .
\end{equation}
Analogously, by substituting Eqs.~(\ref{P(0)}) and
(\ref{propamenosdeC}) into Eq.~(\ref{equivalente2}) we have
that
${\partial C(x'',x')/\partial x'}=0$.
The last two relations tell us that $C(x'',x')=C$, that is,
it is a constant independent of $x''$ and $x'$. 
In order to determine the value of $C$, we first take the limit 
$\tau\rightarrow 0^{+}$ on 
$\langle x'',\tau|x',0\rangle$. 
If we use Eq.~(\ref{propamenosdeC}), we find that:
\begin{equation}
\lim_{\tau\rightarrow 0^{+}}\langle x'',\tau|x',0\rangle
=\lim_{\tau\rightarrow 0^{+}}\frac{C}{\sqrt{\omega\tau}}
\exp\bigl[\frac{im}{2\hbar\tau}(x''-x')^{2}\bigr] 
= C\sqrt{\frac{2\pi i\hbar}{m\omega}}
\delta(x''- x')\,.
\end{equation}
If we compare this result with the initial condition, 
Eq.~(\ref{condicaoinicialSchwing}), we obtain 
$C=\sqrt{m\omega/2\pi i\hbar}$. 
By substituting this result for $C$ into 
Eq.~(\ref{propamenosdeC}), we obtain the desired 
Feynman propagator for the harmonic oscillator:
\begin{eqnarray}
\label{propagadorschwinger}
K(x'',x';\tau)
&=&\langle x'',\tau|x',0\rangle\cr\cr
&=&\sqrt{\frac{m\omega}{2\pi i\hbar\sin(\omega\tau)}}
\,\exp\biggl\{\frac{im\omega}{2\hbar\sin(\omega\tau)}
\bigl[({x''}^{2}+{x'}^{2})
\cos(\omega\tau)-2x''x'\bigr]\biggr\}\, .
\end{eqnarray}
In Sec.~\ref{eigenvalues} we shall see how to extract from 
Eq.~(\ref{propagadorschwinger}) the eigenfunctions and 
energy eigenvalues for the harmonic oscillator and also how 
to obtain, starting from the Feynman propagator, the 
corresponding partition function.

For other applications of this method we 
suggest the following problems for the interested reader. 
Calculate the Feynman propagator using Schwinger's method for
(i) the constant force problem; (ii) a charged spinless particle 
in a uniform magnetic field; and (iii) a charged spinless particle 
in a harmonic oscillator potential placed in a uniform magnetic
field. 

We finish this section by mentioning that Schwinger's method 
can be applied to time-dependent Hamiltonians as
well.\cite{Horing86,Farina93} It also provides a natural
way of establishing the midpoint rule in the path integral
formalism (see Sec.~\ref{pathintegral}) when electromagnetic fields
are present.\cite{FarinaRabello95}

\section{Algebraic Method}\label{algebra}
The origin of the algebraic method dates back to the beginning 
of quantum mechanics, with the matrix formulation of Jordan, 
Heisenberg, and Pauli among others. 
Here, we present an algebraic method for calculating Feynman 
propagators which involves manipulations of momentum and 
position operators.\cite{Wilcox,Barut,Wybourne,Gilmore,Truax} 
This is a powerful method because it is connected with the
dynamical symmetry groups of the system at hand. 
A knowledge of the underlying Lie algebra 
can be used to calculate eigenvalues without explicit knowledge 
of the eigenfunctions.\cite{Yonei,Barut,Wybourne}
It can also be used to calculate propagators for a wide range 
of problems.\cite{Wang,MS82,VB,BV,BSV,DB9194}
A coherent-state version of the algebraic method for 
different problems has been discussed also.\cite{VB,BV89}
Because the use of these mathematical tools can be a bit
cumbersome at first reading, we prefer to explore a 
simpler version of this method, which is close to that 
in Ref.~\onlinecite{Wang}. 
For this purpose, the calculation of the propagator for the
one-dimensional harmonic oscillator is excellent. 

The Hamiltonian operator 
$\hat H$ for a non-relativistic system can usually be written as
a sum of terms involving the operators 
${\hat P}$ and ${\hat X}$ which do not commute. 
Hence, the factorization of the time evolution operator 
$\hat U(\tau)=\exp(-i\tau \hat H/\hbar)$
into a product of simpler exponential\break operators involves
some algebra. 
This algebra deals basically with the commutation relations 
among these non-commuting operators, and uses formulae generically
known as Baker-Campbell-Hausdorff (see Eq.~(\ref{BCH})).
The use of those formulae is the essence of the algebraic method,
because it is easier to calculate the action of these simpler
exponential operators on the states $|x \rb$ or $|p \rb$, than to
calculate the action on these same states of the original time
evolution operator. The algebraic method can be
summarized by the following steps:

\begin{enumerate}
\item First rewrite the evolution operator $\hat
U(\tau)$ as a product of exponentials of the operators ${\hat X}$,
${\hat P}$, and ${\hat P}{\hat X}$. 
(Note that, in contrast with Schwinger's method, 
here the operators $\hat P$ and $\hat X$ are time independent, 
that is, they are in the Schr\"odinger representation.) 
The factorization can be done with the help of the
Baker-Campbell-Hausdorff formula\cite{Wilcox,Merzbacher} 
\begin{equation}
\label{BCH}
e^{A}Be^{-A}=C,
\end{equation}
where $A$, $B$, and $C$ are operators (for simplicity we 
omit the hat on the operators) and
\begin{equation}
\label{CAB}
C=B+[A,B]+\frac{1}{2!}[A,[A,B]]+\frac{1}{3!}[A,[A,[A,B]]]+ \ldots ,
\end{equation}
valid for any $A$ and $B$. Equation~(\ref{CAB}) can be iterated
as: 
\begin{eqnarray}
C^{2}&=&(e^{A}Be^{-A})(e^{A}Be^{-A})=e^{A}B^{2}e^{-A} \nonumber
\\
&\vdots& \nonumber\\
C^{n}&=& e^{A}B^{n}e^{-A} \label{above}
\end{eqnarray}
If we expand $\exp(C)$ and identify each power $C^n$ in
Eq.~(\ref{above}), we find
\begin{equation}
\label{BCHexp}
e^{C}=e^{A}e^{B}e^{-A}\,, 
\end{equation}
which can be inverted to give:
\begin{equation}
\label{imp1}
e^{B}=e^{-A}e^{C}e^{A}\,.
\end{equation}
We then identify $B=-i\tau \hat H/\hbar$ and find a
factorized form of the evolution operator for a conveniently
chosen operator 
$A$. The specific choice for $A$ depends on the explicit form 
of the Hamiltonian. 
This factorization can be repeated as many times as needed. 
Note that, in general, the operator $C$ in Eq.~(\ref{CAB}), 
which is an infinite series with $B$ and multiple commutators 
of $A$ and $B$, is more complicated than the operator $B$ alone, 
which is proportional to the Hamiltonian. 
However, if we choose the operator $A$ conveniently, this series 
can terminate and the remaining terms from the commutators can 
cancel some of those terms originally present in $B$. A more
systematic way of doing this factorization is to use the Lie
algebra related to the problem under study. This way will be
sketched at the end of this section.

\item Next substitute the factorized Hamiltonian into the
definition of the Feynman propagator $K(x'',x';\tau)$, 
Eq.~(\ref{propagadorFeynman}), and calculate the action of the 
exponential of the operators ${\hat X}$, ${\hat P}$, and 
${\hat P}{\hat X}$ on the state $|x \rb$. For the operator 
${\hat X}$ this calculation is trivial and for ${\hat P}$ we just
need to use the closure relation $\uma=\! \int \! dp|p \rb \lb
p|$ and the matrix element 
$\lb x|p \rb =(1/2\pi\hbar)^{1/2}\exp(ixp)$. 
For the mixed operator ${\hat P}{\hat X}$ we need to use the 
expression:
\begin{equation}
\label{p'p}
\lb p'|\exp (-i \gamma \hat P \hat
X/\hbar)|p\rb = e^{-\gamma} \delta(p'-e^{-\gamma}p)\,,
\end{equation}
where $\gamma$ is an arbitrary parameter to be chosen later. 
Equation~(\ref{p'p}) comes from the relation 
\begin{equation}
\label{expPXstate}
\exp(-i\gamma{\hat P}\hat
X/\hbar)|p \rb = e^{-\gamma}|e^{-\gamma}p \> \,.
\end{equation}
\end{enumerate}

Before we apply the algebraic method to a specific problem, we
derive Eq.~(\ref{expPXstate}). We first note
that
\begin{eqnarray}
\label{propintermalg1}
\exp (-i \gamma \hat P \hat X/\hbar){\hat P}|p\rb
= p\exp (-i\gamma{\hat P}{\hat X}/\hbar)|p\rb
\,.
\end{eqnarray}
Equation~(\ref{propintermalg1}) can be rewritten as 
\begin{eqnarray}
\label{expPexp}
\bigl[\exp (-i \gamma{\hat P}{\hat X}/\hbar)
{\hat P}\exp(i \gamma{\hat P}\hat
X/\hbar)\bigr]\exp(-i \gamma{\hat P}\hat
X/\hbar)|p \rb
=p\exp (-i
\gamma{\hat P}{\hat X}/\hbar)|p\rb\,,
\end{eqnarray}
so that we can use the Baker-Campbell-Hausdorff formulae
(\ref{BCH}) to rewrite the term in the square brackets as:
\begin{eqnarray}
\exp(-i \gamma{\hat P}{\hat X}/\hbar) {\hat P}
\exp(i\gamma{\hat P}{\hat X}/\hbar) &=&
\bigl(1+\gamma+\frac{1}{2!}\gamma^{2}+\frac{1}{3!}\gamma^{3}
+ \ldots\bigr){\hat P} \cr\cr &=& e^{\gamma}{\hat P} \,.
\end{eqnarray}
If we substitute the above result into Eq.~(\ref{expPexp}), we
have:
\begin{equation}
{\hat P}\bigl[\exp(-i \gamma{\hat P}\hat
X/\hbar)|p \rb\bigr]=
e^{-\gamma}p\bigr[\exp(-i \gamma\hat P{\hat
X}/\hbar)|p \rb \bigr],
\end{equation}
which shows that 
$\exp(-i\gamma{\hat P}{\hat X} /\hbar)|p\rb$ 
is an eigenstate of the operator ${\hat P}$ with eigenvalue 
$pe^{-\gamma}$. This eigenstate can be written as 
$|e^{-\gamma}p \rb$, up to a constant $C_\gamma$, so that
\begin{equation}
\exp (-i \gamma{\hat P}\hat
X/\hbar)|p \rb = C_{\gamma}|e^{-\gamma}p \rb \,.
\end{equation}
To determine the constant $C_{\gamma}$, we note that:
\begin{eqnarray}
\lb p'|\exp (\frac{i}{\hbar}\gamma{\hat X}\hat
P)\exp (-i\gamma{\hat P}\hat
X/\hbar)|p \rb = |C_{\gamma}|^{2} \lb e^{-\gamma}p'|e^{-\gamma}p
\rb .
\end{eqnarray}
If we use the relation $[{\hat X}{\hat P},{\hat P}{\hat X}]=0$, 
and Eq.~(\ref{imp1}), we have:
\begin{eqnarray}
\lb p'|\exp ( i \gamma[{\hat X},{\hat P}]/\hbar)|p
\rb = |C_{\gamma}|^{2} \delta (e^{-\gamma}(p'-p)),
\end{eqnarray}
so that
\begin{eqnarray}
\label{cg}
e^{-\gamma}\, \delta(p-p')= |C_{\gamma}|^{2}\, e^{\gamma}\,
\delta(p'-p).
\end{eqnarray}
Equation~(\ref{cg}) determines $C_{\gamma}=e^{-\gamma}$ and
finally yields Eq.~(\ref{expPXstate}). 

Now we are ready to apply the algebraic method to solve 
some quantum mechanical problems. 
For the harmonic oscillator the time evolution operator 
(\ref{evolution}) becomes:
\begin{equation}
\label{OPevolucao}
\hat U(\tau)=\exp\bigl[-i
\tau\bigl(\frac{\hat{P}^{2}}{2m}+\frac{1}{2}
m\omega^{2}\hat{X}^{2}\bigr)/\hbar\bigr]\,.
\end{equation}
We follow step (1), choose $A=\alpha{\hat X}^{2}$ where 
$\alpha$ is an arbitrary parameter and $B=-i\tau \hat H/\hbar$, 
and obtain from Eqs.~(\ref{CAB}) and (\ref{BCHexp})
\begin{eqnarray}
&&\exp(\alpha{\hat X}^{2})\exp(-i\tau \hat H/\hbar)
\exp(-\alpha{\hat X}^{2}) \cr\cr
&&=\exp\left\{\frac{-i\tau}{\hbar}
\left[\frac{{\hat P}^{2}}{2m}
+\frac{i\hbar\alpha}{m}({\hat X}{\hat P}+{\hat P}{\hat X})
+\frac{m}{2}\biggl[\omega^{2}
-\biggl(\frac{2\alpha\hbar}{m}\biggr)^{2}
\biggr]{\hat X}^{2}\right]\right\}. \label{aboveeq}
\end{eqnarray} Note that even though the Baker-Campbell-Hausdorff
formulae have an infinite number of terms, the number of
non-vanishing commutators between 
$\hat H$ and ${\hat X}^2$ is finite. 

The general idea of the algebraic method is that we want to
factorize the time evolution operator. In this
case, a step in this direction corresponds to canceling the
term with
${\hat X}^2$ in the right-hand side of Eq.~(\ref{aboveeq}). This is
easily achieved with the choice
\begin{equation}
\label{defalpha}
\alpha=\frac{m\omega}{2\hbar}\,.
\end{equation}
Then by using the commutation relation 
$[{\hat X},{\hat P}]=i\hbar$, we have:
\begin{equation}
\label{OPevolucao2}
\exp\bigl(- i \tau\hat
H/\hbar\bigr) =e^{i\omega\tau/2}\exp(-\alpha\hat
X^{2})\exp\biggl[-\frac{i}{\hbar}\tau\bigl(\frac{{\hat P}^{2}}{2m}
+i\omega{\hat P}{\hat X}\bigr)\biggr]\exp(\alpha{\hat X}^{2})\,.
\end{equation}

We can repeat step (1) to reduce the above operator 
containing ${\hat P}^2$ between brackets into a product of simpler
terms. This time we need to use products of ${\hat P}^2$ instead
of 
${\hat X}^2$. If we use Eqs.~(\ref{BCH}) and (\ref{CAB}), we
obtain:
\begin{equation}
\label{Uintermediario}
\exp(\beta{\hat P}^{2})\bigl(\frac{{\hat P}^{2}}{2m}
+i\omega{\hat P}{\hat X}\bigr)\exp(-\beta{\hat P}^{2})
=\bigl(\frac{1}{2m}+2\omega\hbar\beta\bigr){\hat P}^{2}
+i\omega{\hat P}{\hat X}
\end{equation}
and to eliminate the term proportional to ${\hat P}^{2}$ on the 
right-hand side we take:
\begin{equation}
\label{defbeta}
\beta=-\frac{1}{4m\omega\hbar}\,,
\end{equation}
which gives from Eqs.~(\ref{imp1}) and (\ref{Uintermediario})
\begin{equation}
\label{sep2}
\exp\bigl[-\frac{i}{\hbar}\tau\bigl(\frac{{\hat P}^{2}}{2m}
+i\omega{\hat P}{\hat X}\bigr)\bigr]
=\exp(-\beta{\hat P}^{2})
\exp\bigl[-\frac{i}{\hbar}(i\omega\tau){\hat P}{\hat X}\bigr]
\exp(\beta{\hat P}^{2})\,.
\end{equation}
If we substitute Eq.~(\ref{sep2}) into Eq.~(\ref{OPevolucao2}), 
we have that:
\begin{equation}
\exp\biggl(-\frac{i}{\hbar}\tau\hat H\biggr)
=e^{i\omega\tau/2}\exp(-\alpha{\hat X}^{2})
\exp(-\beta{\hat P}^{2})
\exp\biggl[-\frac{i}{\hbar}(i\omega\tau){\hat P}{\hat X}\biggr]
\exp(\beta{\hat P}^{2})\exp(\alpha{\hat X}^{2})\,. \label{this}
\end{equation}
Equation~(\ref{this}) is the expression for the time evolution
operator written as a product of simpler operators obtained by
applying step (1) of the algebraic method. 

We next follow step (2), insert Eq.~(\ref{this}) into the 
definition of the Feynman propagator 
Eq.~(\ref{propagadorFeynman}), and find
\begin{eqnarray}
\label{propalginterm1}
&& K(x'',x';\tau)=\exp\bigl[-\alpha({x''}^{2}-{x'}^{2})
 +\frac{i\omega\tau}{2}\bigr]\cr\cr
\quad &&\times \!\int\! \frac{dpdp'}{2\pi\hbar}
\exp\bigl[\frac{i}{\hbar}(p'x''-px')
-\beta({p'}^{2}-p^{2})\bigr]\lb
p'|\exp\bigl[-\frac{i}{\hbar}(i\omega\tau) {\hat P}{\hat X}\bigr]|p
\rb \,.\quad
\end{eqnarray}

If we use Eq.~(\ref{p'p}) with $\gamma=i\omega\tau$, and
the definitions (\ref{defalpha}) and (\ref{defbeta}),
we have:
\begin{eqnarray}
K(x'',x';\tau)&=&\frac{1}{2\pi\hbar}
\exp\Biggl[-\frac{m\omega}{\hbar}
\Biggl(({x''}^{2}-{x'}^{2})
+2\frac{(e^{-i\omega\tau}x''-x')^{2}}{1-e^{-2i\omega\tau}}
\Biggl)-\frac{i\omega\tau}{2}\Biggr]\cr\cr
&\times&\int \! 
dp\exp{\Biggl[-\biggl(\frac{1-e^{-2i\omega\tau}}{4m\omega\hbar}\biggr)
\biggl(p-2im\omega
\frac{(e^{-i\omega\tau}x''-x')}{1-e^{-2i\omega\tau}}\biggr)^{2}\Biggr]}.
\end{eqnarray}
This integral has a Gaussian form and can be easily done 
giving the harmonic oscillator propagator:
\begin{equation}
K(x'',x';\tau)=\sqrt{\frac{m\omega}{2\pi i\hbar\sin\omega\tau}}
\exp\biggl[\frac{im\omega}{2\hbar\sin\omega\tau}
\biggl(({x''}^{2}+{x'}^{2})
\cos\omega\tau-2x''x'\biggr)
\biggr],
\end{equation}
where we used that 
$(1-e^{-2i\omega\tau})=2ie^{-i\omega\tau}\sin\omega\tau$ 
and Euler's formula, $e^{i\omega\tau}=\cos\omega\tau
+i\sin\omega\tau$.
This result naturally agrees with the one obtained in 
Sec.~\ref{algebra} using Schwinger's method.

Before we finish this section, we want to comment that the 
algebraic method can be discussed on more formal grounds, 
identifying the underlying Lie algebra, and using it to 
explicitly solve the problem of interest. 
For the one-dimensional harmonic oscillator we can find a 
set of operators
\begin{eqnarray}
L_-=-\frac 12 \partial_{xx}, \qquad
L_+=\frac 12 x^2, \qquad
L_3=\frac 12 x\partial_x + \frac 14 \,,
\end{eqnarray}
such that the Hamiltonian operator can be written as 
$\hat H=(\hbar/m)L_{-}+m\omega^2L_+$.
The above operators satisfy the SO(3) Lie algebra
\begin{eqnarray}
[L_+,L_-]=2L_3, \qquad
[L_3,L_{\pm}]=\pm L_{\pm} \,.
\end{eqnarray}
This algebra is isomorphic to the usual SU(2) Lie algebra of the 
angular momenta and can be used to construct specific
Baker-Campbell-Hausdorff formulae,\cite{Gilmore,Truax} so that
given the algebra, the solution arises naturally.\cite{Wang} If
one considers three-dimensional problems, which are more
involved because of the presence of terms proportional
to $1/r^2$, this algebra can still be used to find the propagator,
but the operators have to be generalized.\cite{Barut,Wybourne} 
These generalized operators can be used to solve a
wide range of problems.\cite{MS82,VB,BV,BSV,DB9194}

\section{Path Integral Method}\label{pathintegral}
The path integral formalism was introduced by
Feynman\cite{Feynman48} in 1948, following earlier ideas developed
by Dirac.\cite{Dirac33} In the last few decades, path integral
methods have become very popular, mainly in the
context of quantum mechanics, statistical physics, and quantum
field theory. Since the pioneering textbook of Feynman and
Hibbs,\cite{FeynmanHibbs} many others have been written on this
subject, not only in quantum 
mechanics,\cite{Inomata,Schulman,Khandekar,Kleinert,GroscheSteiner} 
but also in condensed matter,\cite{Amit} as well as quantum field 
theory,\cite{Fried,Rivers,Das} to mention just a few.

In this section we shall apply Feynman's method to again obtain the 
harmonic oscillator quantum propagator already established in 
Secs.~\ref{schwinger} and \ref{algebra}. The purpose here is to 
evaluate the
corresponding path integral explicitly, without making use of the
semiclassical approach which
is often used in the literature.  Of course
this kind of direct calculation already exists in the literature,
see for example Refs.~\onlinecite{Holstein98},
\onlinecite{Das}, and \onlinecite{Itzyckson}. However, we shall
present an alternative and very simple procedure.

The path integral expression for the quantum propagator is
formally given by:
\begin{equation}\label{propagFeynmanformal}
K(x_N,x_0;\tau)=\!\int_{
{\,}^{x(0)=x_0}_{x(\tau)=x_N}
}[Dx] 
e^{iS(x)/\hbar}\, ,
\end{equation}
where $S(x)$ is the action functional:
\begin{equation}
\label{acao}
S(x) \equiv \! \int_{t_{0}}^{t_{N}} \bigl[\frac 12m\dot x^{2}(t)
-V(x(t))\bigr]dt,
\end{equation}
and $[Dx]$ is the functional measure. 
According to Feynman's prescription, we have that:
\begin{eqnarray}
\label{FeynmansPresc}
K(x_{N},x_{0};\tau)
&=&\lim_{{\;}^{N\rightarrow\infty}_{\;\;\varepsilon\rightarrow 0}}
{\sqrt{\frac{m}{2\pi i\hbar\varepsilon}}}\int\prod_{j=1}^{N-1}
\biggl({\sqrt{\frac{m}{2\pi i\hbar\varepsilon}}dx_{j}}\biggr)
\nonumber\\
\nonumber\\
&&\times{}
\exp\Biggl\{\frac{i}{\hbar}\sum_{k=1}^{N}
\Biggl[\frac{m(x_{k}-x_{k-1})^{2}}{2\varepsilon}
-\varepsilon V\Biggl(\frac{x_{k}+x_{k-1}}{2}\Biggr)\Biggr]\Biggr\}\, ,
\end{eqnarray}
where $N\varepsilon=\tau$. With this prescription, the scenario 
is the following: summation over all the functions $x$ 
means to sum over all the polynomials in the plane ($t,x(t)$), 
starting at 
$(x_{0},t_{0})$ and finishing at $(x_{N},t_{N})$, which gives 
rise to the integrations over the variables $x_{j} \equiv
x(t_{j})$ from $-\infty$ to $\infty$, where
$t_j=t_0+j\varepsilon$, with 
$j=1, 2, \dots, N-1$. 
Hence, to evaluate a path integral means to calculate an
infinite number of ordinary integrals, which requires some kind
of recurrence relation. 

When electromagnetic potentials are absent as is the case
here, it is not necessary to
adopt the midpoint rule for the potential $V(x)$ as given by 
Eq.~(\ref{FeynmansPresc}), and other choices can also be made. 
Instead of using the midpoint rule we shall 
write the discretized version of the action as
\begin{equation}
S\cong\sum_{j=1}^{N}\frac{m(x_{j}-x_{j-1})^{2}}{2\tau_{j}}-\tau_{j}
\frac 12\bigl(V(x_{j})+V(x_{j-1})\bigr)\,,
\end{equation}
where for generality we have taken $\tau_j$ as the $j$th time
interval so that 
$\tau=\sum\!\!\!\!\!\!\!\!\!
\!_{{\;}_{{\;}_{{\;}_{j=1}}}}^{{\;}^{{\;}^{\;\; N}}}\tau_j$.
%
%
%\!\!\!\!\!\!\!_{\;\;\atop{j=1}}^{{\;}^{{\;}^N}}\tau_{j}$. 
%
%
%
Then, we write the Feynman propagator in the form:
\begin{equation}
\label{propagadorintpropagadorinf}
K(x_{N},x_{0};\tau)=\lim_{N\rightarrow\infty}\int\prod_{j=1}^{N}
{K(x_{j},x_{j-1};\tau_{j})}\prod_{k=1}^{N-1}dx_{k}\,,
\end{equation}
where the propagator for an infinitesimal time interval is given
by:
\begin{equation}
\label{propagadorinfinitesimal1}
K(x_{j},x_{j-1};\tau_{j})=\sqrt{\frac{m}{2\pi i\hbar\tau_{j}}}
\exp{\biggl\{\frac{i}{\hbar}
\biggl[\frac{m(x_{j}-x_{j-1})^{2}}{2\tau_{j}}-\tau_{j}\frac 12
\biggl(V(x_{j})+V(x_{j-1})\biggr)\biggr]\biggr\}}\, .
\end{equation}

If we use the Lagrangian for the harmonic oscillator, namely,
\begin{equation}
L(x,\dot x)=\frac 12m\dot x^{2}-\frac 12m\omega^{2}x^{2}\, ,
\end{equation}
the infinitesimal propagator, given by 
Eq.~(\ref{propagadorinfinitesimal1}), takes the form:
\begin{equation}
\label{propagadorinfinitesimal2}
K(x_{j},x_{j-1};\tau_{j})=\sqrt{\frac{m\omega}{2\pi\hbar}}
\sqrt{\frac{1}{\omega\tau_{j}}}
\exp\left\{\frac{im\omega}{2\hbar}
\frac{1}{\omega\tau_{j}}\biggl[
\biggl(1-\frac{\omega^{2}\tau_{j}^{2}}{2}\biggr)
(x_{j}^{2}+x_{j-1}^{2})\biggr]-2x_{j}x_{j-1}\right\}
\end{equation}

To calculate this infinitesimal propagator we now
define new variables $\phi_j$ such that:
\begin{equation}
\label{defphi}
\sin\phi_{j}=\omega\tau_{j},
\end{equation}
which implies that $\phi_{j}\cong\omega\tau_{j}$ and
$\cos\phi_{j} \cong 1
-\omega^{2}\tau_{j}^{2}/2$.
In fact, other variable transformations could be
tried, but this is the simplest one that we were able to find that
allows an easy iteration through a convolution-like formula. It is
also helpful to introduce a function $F$:
\begin{equation}
\label{defF}
F(\eta,\eta';\phi)=\sqrt{\frac{m\omega}{2\pi i\hbar}}
\sqrt{\frac{1}{\sin\phi}}
 \exp\biggl[\frac{im\omega}{2\hbar}
\frac{1}{\sin\phi}
\bigl(\cos\phi(\eta^{2}+{\eta^{\prime}}^{2})
-2\eta\,\eta^{\;\prime}\bigr)\biggr].
\end{equation}
Then, using Eqs.~(\ref{propagadorinfinitesimal2}), 
(\ref{defphi}), and (\ref{defF}), we can rewrite the harmonic 
oscillator propagator (\ref{propagadorintpropagadorinf}) as:
\begin{equation}
\label{propagadorintF}
K(x_{N},x_{0};\tau)
=\lim_{N\rightarrow\infty}\! \int\prod_{j=1}^{N} \!
{F(x_{j},x_{j-1};\phi_{j})}\prod_{k=1}^{N-1}dx_{k} .
\end{equation}

The function $F$ has an interesting property:
\begin{equation}
\label{propF}
\int_{-\infty}^{\infty}F(\eta'',\eta;\phi'')
F(\eta,\eta';\eta')d\eta
=F(\eta'',\eta';\phi''+\phi').
\end{equation}
This can be seen by a simple direct calculation. 
>From its definition (\ref{defF}), we have that:
\begin{eqnarray}\label{Fresnel1}
\int_{-\infty}^{\infty}F(\eta'',\eta;\phi'')
F(\eta,\eta';\phi')d\eta
&=&\frac{m\omega}{2\pi i\hbar}\sqrt{\frac{1}{\sin\phi''\sin\phi'}}
\exp\biggl[\frac{im\omega}{2\hbar}
\biggl(\frac{\cos\phi''}{\sin\phi''}{\eta''}^{2}
+\frac{\cos\phi'}{\sin\phi'}{\eta'}^{2}\biggr)\biggr]
\cr\cr\cr
&& {\times} 
\int_{-\infty}^{\infty}d\eta
\exp{\biggl[\frac{im\omega}{2\hbar}(\alpha\,\eta^{2}-2\eta\beta)\biggr]}\,,
\end{eqnarray}
where we have defined:
\begin{eqnarray}
\label{defalphabeta}
\alpha=
\frac{\sin(\phi''+\phi')}{\sin\phi''\sin\phi'} \qquad \mbox{and}
\qquad
\beta=
\frac{\eta''\sin\phi'
+\eta^{\,\prime}\sin\phi''}{\sin\phi''\sin\phi'}\,.
\end{eqnarray}

By completing the square in the integrand of Eq.~(\ref{Fresnel1}) 
and calculating the Fresnel integral, we obtain:
\begin{eqnarray}\label{Fresnel2}
\int_{-\infty}^{\infty}
F(\eta'',\eta;\phi'')F(\eta,\eta';\phi')d\eta&=&
\frac{m\omega}{2\pi i\hbar}
\sqrt{\frac{2\pi i\hbar}{m\omega\alpha}}
\sqrt{\frac{1}{\sin\phi''\sin\phi'}}\cr\cr
&& {\times}
\exp\biggl[\frac{im\omega}{2\hbar}\biggl(\frac{\cos\phi''}{\sin\phi''}
{\eta''}^{2}
+\frac{\cos\phi'}{\sin\phi'}{\eta'}^{2}
-\frac{\beta^{2}}{\alpha}\biggr)\biggr]\,.
\end{eqnarray}
If we use the definitions in Eq.~(\ref{defalphabeta}), 
as well as some trivial manipulations with trigonometric 
functions, it is straightforward to show that:
\begin{eqnarray}
&&\int_{-\infty}^{\infty}F(\eta'',\eta;\phi'')
F(\eta,\eta';\phi')d\eta\cr\cr
&& =\sqrt{\frac{m\omega}{2\pi i\hbar}}
\sqrt{\frac{1}{\sin(\phi''+\phi')}}
\exp\bigl[\frac{im\omega}{2\hbar}
\frac{1}{\sin(\phi''+\phi')}\bigl(\cos(\phi''+\phi')({\eta''}^{2}
+{\eta^{\,\prime}}^{2})-2\eta''\eta^{\,\prime}\bigr)\bigr],
\end{eqnarray}
which is precisely Eq.~(\ref{propF}). We now define $x''=x_N$,
$x'=x_0$, use the fact that 

$\lim\!\!\!\!\!\!\!\!\!\!\!_{{\;}_{{\;}_{N\to\infty}}}
\sum\!\!\!\!\!\!\!\!\!
\!_{{\;}_{{\;}_{{\;}_{j=1}}}}^{{\;}^{{\;}^{\;\; N}}}\phi_j
%
%\sum\!\!\!\!\!\!\!_{\;\;\atop{j=1}}^{{\;}^{{\;}^N}}\phi_{j}
%
=\omega\tau$ and the result in 
Eq.~(\ref{propagadorintF}) to finally obtain the desired 
propagator:
\begin{eqnarray}
K(x'',x';\tau)&=&F(x'',x';\omega\tau)\cr\cr
%\sum_{j=0}^{N}\phi_{j}
&=&\sqrt{\frac{m\omega}{2\pi i\hbar\,\sin\omega\tau}}
\exp\biggl[\frac{im\omega}{2\hbar\,\sin\omega\tau}
\biggl(\cos\omega\tau({x''}^{2}+{x'}^{2})-2x''x'\biggr)
\biggr]\,.
\label{harmoscprop}
\end{eqnarray}
This result coincides with the expressions obtained in the
previous sections for the harmonic oscillator. 
Let us now review some of the applications of the Feynman
propagator. 

\section{Eigenfunctions, Eigenvalues and the 
Partition Function}\label{eigenvalues}
In this section we will
show how to obtain the stationary states and the corresponding
energy levels, as well as the partition function of the quantum
harmonic oscillator, directly from the expression of the Feynman
propagator calculated earlier. Although these tasks are well known
in the literature (see for instance Refs.~\onlinecite{Das}
and \onlinecite{DH88}), we shall present them here for
completeness.

To obtain the energy eigenstates and eigenvalues, we
need to recast the propagator (\ref{harmoscprop}) in a form that
permits a direct comparison with the spectral 
representation for the Feynman propagator given by:
\begin{equation}
\label{espectral}
K(x,x';\tau)=\Theta(\tau)\! \sum_n\phi_n(x)\phi_{n}^{*}(x')
e^{-iE_n\tau/\hbar}\,.
\qquad (\tau>0)
\end{equation}
If we define the variable $z=e^{- i\omega \tau}$, 
we can write:
\begin{subequations}
\begin{eqnarray}
\sin(\omega\tau)&=&\frac{1}{2i}\frac{1-z^{2}}{z} \\
\cos(\omega\tau)&=&\frac{1+z^{2}}{2z}\,.
\end{eqnarray}
\end{subequations}
We further define $\xi' \equiv \sqrt{m\omega/\hbar}x'$ and 
$\xi''\equiv \sqrt{m\omega/\hbar}x''$, and express the harmonic
oscillator propagator to the form:
\begin{eqnarray}
\label{propMehler}
K(x'',x';\tau)&=&\sqrt{\frac{m\omega z}{\pi\hbar}}
(1-z^{2})^{-1/2}
\exp{\biggl\{\frac{1}{1-z^{2}}\bigl[2\xi'\xi'' z
-({\xi'}^{2} + {\xi''}^{2}) 
\bigl(\frac{1+z^{2}}{2}\bigr)\bigr]\biggr\}} \\
&=&\sqrt{\frac{m\omega z}{\pi\hbar}}(1-z^{2})^{-1/2}
\exp{\bigl[-\frac 12({\xi'}^{2}+{\xi''}^{2})\bigr]}
\exp\bigl[\frac{2\xi'\xi'' z -({\xi'}^{2}+{\xi''}^{2})z^{2}}{1-z^{2}}
\bigr]\,,\cr
&{\;}&
\end{eqnarray}
where we used the identity
\begin{eqnarray}
\frac{1+z^{2}}{2(1-z^{2})}=\frac 12
+\frac{z^{2}}{1-z^{2}}\, .
\end{eqnarray}

Now we consider Mehler's formula:\cite{Arfken}
\begin{equation}
\label{Mehler}
(1-z)^{-1/2}
\exp{\biggl[\frac{2xyz-(x^{2}+y^{2})z^{2}}{1-z^{2}}
\biggr]}
=\sum_{n=0}^{\infty}
{H_{n}(x)H_{n}(y)\frac{z^{n}}{2^{n} n!}} .
\qquad (|z|<1)
\end{equation}
However, some care must be taken in order to use Eq.~(\ref{Mehler})
in Eq.~(\ref{propMehler}), because $|z|=1$ and Mehler's formula 
(\ref{Mehler}) requires that $|z|<1$. 
This problem can be circumvented if we add an imaginary 
part to $\omega$, namely, if we let 
$\omega\rightarrow\omega - i\varepsilon$, 
and take $\varepsilon\rightarrow 0$ after the calculations. 
Hence, if we use Eq.~(\ref{Mehler}), Eq.~(\ref{propMehler}) 
takes the form:
\begin{equation}
\label{propagadorHnHn}
K(x'',x';\tau)=\sqrt{\frac{m\omega}{\pi\hbar}}
\exp{\bigl[-\frac{m\omega}{2\hbar}({x''}^{2}+{x'}^{2})\bigr]}
\sum_{n=0}^{\infty} H_{n}
\bigl(\sqrt{\frac{m\omega}{\hbar}}x''\bigr)H_{n}
\bigl(\sqrt{\frac{m\omega}{\hbar}}x'\bigr)
\frac{e^{-i\omega\tau (n+1/2)}}{2^{n} n!},
\end{equation}
where we have let
$x=\xi''=\sqrt{m\omega/\hbar}x''$ and
$y=\xi'=\sqrt{m\omega/\hbar}x'$.

If we compare Eq.~(\ref{propagadorHnHn}) with the spectral
representation (\ref{espectral}), we finally obtain the well
known results for the energy eigenfunctions (apart from a phase
factor) and energy levels, respectively:
\begin{equation}
\phi_{n}(x)=\frac{1}{\sqrt{2^{n} n!}}
\bigl(\frac{m\omega}{\pi\hbar}\bigr)^{1/4}
\exp{\bigl(-\frac{m\omega}{\pi\hbar}x^{2}\bigr)}
H_{n}\bigl(\sqrt{\frac{m\omega}{\hbar}}x\bigr)
\end{equation}
\begin{equation}
E_{n}=(n+\frac 12)\hbar\omega\,.
\end{equation}

We finish this section by calculating the partition 
function for the harmonic oscillator. 
With this purpose in mind, recall that the partition function 
in general can be written as:
\begin{eqnarray}
Z(\beta)={\rm Tr}\, e^{-\beta\hat H}\,.
\end{eqnarray}
The trace operation can be taken over a discrete basis, the
eigenfunctions of the Hamiltonian itself,
or, more conveniently here, over the continuous set of
eigenstates of the position operator (denoted by $\vert
x\rangle$):
\begin{eqnarray}
Z(\beta)= \!\int_{-\infty}^{+\infty} dx \, \langle x\vert\, 
e^{-\beta\hat H}\,
\vert x\rangle\,.
\end{eqnarray}
If we identify the integrand with the Feynman propagator with 
the end points $x'=x''=x$ and $\beta=i\hbar\tau$
as the imaginary time interval, we have
\begin{eqnarray}
Z(\beta)=\!\int_{-\infty}^{+\infty}\! dx\, 
K(x,x;-i\hbar\beta)\,.
\label{partitionfunction}
\end{eqnarray}
Then, from the harmonic oscillator propagator
Eq.~(\ref{harmoscprop}), we readily obtain:
\begin{equation}\label{propcombeta}
K(x,x;-i\hbar\beta)=
\sqrt{\frac{m\omega}{2\pi \hbar\sinh (\omega\beta\hbar)}}
\exp\bigl[-\frac{m\omega}{\hbar\sinh(\omega\beta\hbar)}
(\cosh (\omega\beta\hbar)-1)\, x^2\bigr]\,,
\end{equation}
where we have used $\sin (-i\alpha)=-i\sinh\alpha$ and 
$\cos (i\alpha)=\cosh\alpha$. 
By substituting Eq.~(\ref{propcombeta}) into
Eq.~(\ref{partitionfunction}), and evaluating the remaining
Gaussian integral, we finally obtain:
\begin{eqnarray}
Z(\beta)&=&\sqrt{\frac{m\omega}{2\pi\hbar\sinh (\omega\beta\hbar)}}
\int_{-\infty}^{\infty}
\exp\biggl[-\frac{m\omega \, x^2}{\hbar}
\tanh\bigl(\frac{\omega\beta\hbar}{2}\bigr)
\biggr]dx\cr\cr
&=&\frac{1}{2\sinh\bigl(\frac 12\omega\beta\hbar\bigr)}\,,
\label{93}
\end{eqnarray}
where we used the identities $\cosh\alpha -1=2\sinh^{2}(\alpha /2)$ 
and $\sinh (\alpha)=2\sinh (\alpha /2)\cosh(\alpha /2)$. 
Equation~(\ref{93}) is the partition function for the
one-dimensional harmonic oscillator.

\section{Conclusions}
We have rederived the one-dimensional harmonic oscillator 
propagator using three different techniques. First we used a 
method developed by Schwinger that is usually used in quantum
field theory, but that is also well suited for non-relativistic
quantum mechanical problems although rarely used. 
We hope that our presentation of this method will
help it become better known among physics teachers and students. 
Then we presented an algebraic method that deals with the
factorization of the time evolution operator using the
Baker-Campbell-Hausdorff formulae. We hope that our presentation
will motivate teachers and students to learn more about such
powerful methods, which are closely connected with the use of Lie
algebras. It is worth mentioning that these methods 
can be applied not only in
non-relativistic quantum mechanics, but also in the context of
relativistic theories. Finally, we presented a direct calculation
of the Feynman path integral for the one-dimensional harmonic
oscillator using a simple but very convenient recurrence relation
for the composition of infinitesimal propagators. We hope that
the presentation of these three methods together with the
calculation of the Feynman propagator will help readers compare
the advantages and difficulties of each of them. 

\begin{acknowledgments}
The authors would like to acknowledge the
Brazilian agencies CNPq and CAPES for partial financial
support.
\end{acknowledgments}

\end{document}